\documentstyle[12pt]{article}

\newcommand{\EQ}{\begin{equation}}
\newcommand{\EN}{\end{equation}}
\newcommand{\bea}{\begin{eqnarray}}
\newcommand{\ena}{\end{eqnarray}}
\newcommand{\bdis}{\begin{displaymath}}
\newcommand{\edis}{\end{displaymath}}
\newcommand{\vs}[1]{\vspace{#1 mm}}

\renewcommand{\a}{\alpha}

\renewcommand{\d}{\delta}
\renewcommand{\v}{\Delta}

\renewcommand{\t}{\tau}
\newcommand{\ep}{\epsilon}

\newcommand{\pa}{\partial}

\newcommand{\nn}{\nonumber \\}

\begin{document}

\topmargin 0pt
\oddsidemargin 5mm

\newcommand{\NP}[1]{Nucl.\ Phys.\ {\bf #1}}
\newcommand{\PL}[1]{Phys.\ Lett.\ {\bf #1}}
\newcommand{\CMP}[1]{Comm.\ Math.\ Phys.\ {\bf #1}}
\newcommand{\PR}[1]{Phys.\ Rev.\ {\bf #1}}
\newcommand{\PRL}[1]{Phys.\ Rev.\ Lett.\ {\bf #1}}
\newcommand{\PREP}[1]{Phys.\ Rep.\ {\bf #1}}
\newcommand{\PTP}[1]{Prog.\ Theor.\ Phys.\ {\bf #1}}
\newcommand{\PTPS}[1]{Prog.\ Theor.\ Phys.\ Suppl.\ {\bf #1}}
\newcommand{\NC}[1]{Nuovo.\ Cim.\ {\bf #1}}
\newcommand{\JPSJ}[1]{J.\ Phys.\ Soc.\ Japan\ {\bf #1}}
\newcommand{\MPL}[1]{Mod.\ Phys.\ Lett.\ {\bf #1}}
\newcommand{\IJMP}[1]{Int.\ Jour.\ Mod.\ Phys.\ {\bf #1}}
\newcommand{\AP}[1]{Ann.\ Phys.\ {\bf #1}}
\newcommand{\RMP}[1]{Rev.\ Mod.\ Phys.\ {\bf #1}}
\newcommand{\PMI}[1]{Publ.\ Math.\ IHES\ {\bf #1}}
\newcommand{\JETP}[1]{Sov.\ Phys.\ J.E.T.P.\ {\bf #1}}
\newcommand{\TOP}[1]{Topology\ {\bf #1}}
\newcommand{\AM}[1]{Ann.\ Math.\ {\bf #1}}
\newcommand{\LMP}[1]{Lett.\ Math.\ Phys.\ {\bf #1}}
\newcommand{\CRASP}[1]{C.R.\ Acad.\ Sci.\ Paris\ {\bf #1}}
\newcommand{\JDG}[1]{J.\ Diff.\ Geom.\ {\bf #1}}
\newcommand{\JSP}[1]{J.\ Stat.\ Phys.\ {\bf #1}}

\begin{titlepage}
\setcounter{page}{0}
\begin{flushright}
August 1997\\
\end{flushright}

\vs{8}
\begin{center}
{\Large  Algebraic Structure in Non-Orientable \\
 Open-Closed String Field Theories}

\vs{15}
{\large Naohito\ Nakazawa\footnote{e-mail address:
nakazawa@ifse1.riko.shimane-u.ac.jp}} \\
{\em 
Department of Physics, Faculty of Science and Engineering, Shimane University\\
Matsue 690, Japan}  \\
and \\ 
{\large Daiji\ Ennyu} \\
{\em 
Department of Physics, Hiroshima National College of Maritime Technology \\ 
Toyota-Gun, Hiroshima 725-02, Japan} \\
\end{center}

\vs{8}
\centerline{{\bf{Abstract}}}

We apply stochastic quantization method to
real symmetric matrix-vector models for the second quantization of
non-orientable strings, including both open and closed strings.
The Fokker-Planck
hamiltonian deduces a well-defined non-orientable open-closed string field theory at the double scaling limit of the matrix model. There appears a new algebraic structure in the continuum F-P hamiltonian including a Virasoro algebra and a $SO( r )$ current algebra.

\end{titlepage}
\newpage
\renewcommand{\thefootnote}{\arabic{footnote}}
\setcounter{footnote}{0}

Recently, the matrix model approach to superstrings has been investigated extensively as the non-perturbative definition of superstring theories, especially for type IIA~\cite{IIA} and IIB~\cite{IIB} theories. The explicit construction of superstring field theories may be somehow an analogue of the construction of non-critical string field theories\cite{IK} via the double scaling limit of the matrix models~\cite{DS} which provides not only the basis for the non-perturbative analysis of string theories but also the clear understanding of the constraints realized in the algebraic structure of the string field theoretic hamiltonian, such as the Virasoro and the W-algebra~\cite{WAL}. Among many works on the matrix model construction of non-critical string field theories~\cite{IK}\cite{JR}\cite{IIKMNS}\cite{Wa}\cite{Na}\cite{Kos}\cite{Mo}\cite{AJ},  the algebraic structure in non-critical string field theories has been investigated mainly for orientable closed strings~\cite{IK}\cite{JR}\cite{IIKMNS}, non-orientable closed strings~\cite{Na} and orientable open-closed strings~\cite{Kos}\cite{Mo}.  For orientable 2D surfaces with boundaries, an interesting algebraic structure, $sl (r, {\bf C})\times sl (r, {\bf C})$ chiral current algebra including $SU(r)$ current algebra, has been found in a field theoretic hamiltonian of orientable open-closed strings\cite{AJ}. The observation indicates the relation between the algebraic structure and the Chan-Paton like realization of gauge groups in orientable open strings. Since the type I superstring consists of non-orientable open and closed strings, we are interested in the algebraic structure in the string field theoretic hamiltonian for non-orientable open-closed strings.

In this short note, as a toy model of type I strings, we construct non-critical non-orientable open-closed string field theories. To do this, we apply stochastic quantization method\cite{PW} to real symmetric matrix-vector models. The continuum limit of the string field theory is defined by the double scaling limit of the matrix-vector models. We show that the string field theoretic hamiltonian, which is the continuum limit of the loop space Fokker-Planck hamiltonian in stochastic quantization\cite{JR}\cite{Na}, appears in the form of a linear combination of some new deformation generators of non-orientable strings which satisfy algebraic relations including a  Virasoro algebra and a $SO( r )$ current algebra. 

For orientable 2D surfaces with boundaries, the sum of triangulated surfaces is  reconstructed by the hermitian matrix-vector models\cite{Kaz}. 
In order to introduce the non-orientable open string interaction, we consider real symmetric matrix-vector 
models. For one-matrix case, the most general form of the action is given by  
\bea
S(M, V) 
&=& S(M)_{closed} + S(M, V)_{open}          \ , \nn
S(M)_{closed}
&=& - \sum_{\a = 0} {g_\a \over \a + 2} N^{- \a /2}{\rm tr}M^{\a + 2}  \ , \nn
S(M, V)_{open}
&=& - {1\over 2} \sum_{a = 1}^{r} 
\sum_{\a = 0} \mu^a_\a  N^{- \a /2}V^a_i M^\a_{ij}V^a_j \ , 
\ena
where $M_{ij}$ denotes a $N \times N$ real symmetric matrix. $V^a_i$ is the i-th component of a N-dimensional vector and the index $a$ runs from 1 to $r$. Namely, we have introduced $r$ \lq\lq colours \rq\rq which are assigned to each boundary. 

Let us start with the following Langevin equations for the one matrix model,
\bea
{\v}M_{ij}(\t)
&=&  - {\pa \over\pa M}S(M, V)_{ij}(\t) \v\t + \v\xi_{ij}(\t)        \ , \nn
{\v}V^a_i (\t)
&=& - \lambda^a{\pa\over\pa V}S(M, V)^a_i(\t) \v\t + \v\eta^a_i(\t)      \ .  
\ena
We have introduced the scale parameter $\lambda^a$ in the Langevin equation of vector variables. At the double scaling limit, the parameter $\lambda^a$ defines the time scale for the proper time evolution of the open string end-point along the boundaries with the \lq\lq colour \rq\rq index $a$. 
The stochastic time $\t$ is discretized with the unit time step $\v\t$. We define the time evolution, 
$
M_{ij} ( \t+\v\t ) \equiv M_{ij} ( \t ) + \v M_{ij}( \t ) \ 
$
and 
$
V^a_i ( \t+\v\t ) \equiv V^a_i ( \t ) + \v V^a_i( \t ) \ 
$, 
in terms of Ito's stochastic calculus\cite{I}.
The discretized stochastic time development with $\v\t$ 
corresponds to the
one step deformation in dynamical triangulation in random surfaces with open boundaries.
The correlation of the white noise $\v\xi_{ij}$ and $\v\eta^a_i$ are 
defined by
\bea
<\v\xi_{ij}(\t) \v\xi_{kl}(\t)>_\xi
&=& \v\t \big( \d_{il} \d_{jk} + \d_{ik} \d_{jl} \big)   \ , \nn
<\v\eta^a_i(\t) \v\eta^b_j(\t)>_\eta
&=& 2\lambda^a \v\t \d^{ab}\d_{ij}             \ , 
\ena
where we keep the $\lambda^a$-independence of the $equilibrium$ probability distribution. 

The basic field variables are the closed string variables,
$
\phi_n = {\rm tr}(M^n) N^{-1 - {n\over 2}}   \ , 
$
and the open string variables,
$
\psi_n^{ab} = ( V^a M^n V^b )N^{- 1 - {n\over 2}}  \ . 
$
Following to Ito's calculus, we calculate the time development of these string variables\cite{Na}. We obtain 
\bea
\v\phi_n
&=& \v\t { n\over 2} \big\{ \sum_{k=0}^{n-2}
\phi_k \phi_{n-k-2}
+ (n-1) {1\over N} \phi_{n-2}                 \big\}      \nn 
&+& \v\t\ { n\over 2}\sum_{\a = 0}\sum_{a=1}^{N_0}\a {\mu^a_\a  \over N}\psi^{aa}_{n+\a -2} + \v\t\ n \sum_{\a=0} g_\a \phi_{n+\a}   +   \v \zeta_{n-1}    \ ,   \nn
\v\psi^{ab}_n 
&=& 2\lambda^a \v\t \d^{ab}\phi_n + \v\t {n(n-1)\over 2N} \psi^{ab}_{n-2} + \v\t \sum_{k=0}^{n-2}(k+1)\phi_{n-k-2}\psi^{ab}_k         \nn  
&+& \v\t\ n  \sum_{\a=0} g_\a \psi^{ab}_{n+\a} + {1\over 2}\v\t\sum_{\a=1}\sum_{k=0}^{n-1}\sum_{l=0}^{\a -1}\sum_{c=1}^{N_0} \mu^c_\a \psi^{ac}_{k + l}\psi^{cb}_{n-k-l-2+\a}              \nn 
&+& \v\t \sum_{\a=0} ( \lambda^a\mu^a_\a + \lambda^b\mu^b_\a )  \psi^{ab}_{n+\a} + \v\zeta^{ab}_{n-1} + \v\eta^{ab}_n    \ . \nn
\ena
Here the new noise variables are defined by
\bea
\v\zeta_{n-1}
&\equiv& n {\rm tr}(\v\xi M^{n-1}) N^{-1 - {n\over 2}}     \ , \nn 
\v\zeta^{ab}_{n-1} 
&\equiv& \sum_{k=0}^{n-1} (V^a M^k\v\xi M^{n-k-1}V^b )/ N^{1 + n/2}  \ , \nn 
\v\eta^{ab}_n 
&\equiv& \big\{ (\v\eta^a M^n V^b ) + ( V^a M^n \v\eta^b ) \big \} / N^{1 + n/2}                      \ . \nn
\ena
The correlations of these noise variables are given by
\bea
<\v\zeta_{m-1} (\t) \v\zeta_{n-1} (\t)>_{\xi\eta}
&=& \v\t {2\over N^2} n m < \phi_{m+n-2} (\t) >  \      , \nn
<\v\zeta_{m-1} (\t) \v\zeta^{ab}_{n-1} (\t)>_{\xi\eta}
&=& \v\t {2\over N^2} n m < \psi^{ab}_{m+n-2} (\t) >     \ ,  \nn
<\v\zeta^{ab}_{m-1} (\t) \v\zeta^{cd}_{n-1} (\t)>_{\xi\eta}    
&=& \v\t {1\over N}\sum_{k=0}^{n-1}\sum_{l=0}^{m-1} < \big( \psi^{ac}_{k+l} \psi^{bd}_{m+n-k-l-2} + \psi^{ad}_{k+l} \psi^{bc}_{m+n-k-l-2} \big) >             \ , \nn 
<\v\eta^{ab}_m \v\eta^{cd}_n >_{\xi\eta}
&=& \v\t {2\over N}<  \big( \lambda^a\d^{ac}\psi^{bd}_{m+n} +
\lambda^a\d^{ad}\psi^{bc}_{m+n}   \nn 
&{}& \qquad + \lambda^b\d^{bd}\psi^{ac}_{m+n} + \lambda^b\d^{bc}\psi^{ad}_{m+n} \big)   >   \ . 
\ena
These noise correlations are understood in the sense of Ito's calculus. This means that, for example, the expectation value of $\phi_{m+n-2}(\t)$ in R.H.S. of (6) is defined with
respect to the white noise correlations up to the stochastic time $\t - \v\t$.
We notice that 
$
<\v\zeta_n (\t)>_{\xi\eta} = <\v\zeta^{ab}_n (\t)>_{\xi\eta} 
 = <\v\eta^{ab}_n (\t)>_{\xi\eta} = 0   
$
exactly holds. Especially, from eqs. (4)(6), the $\lambda$-independence of the equilibrium limit deduces a S-D equation which ensures the $SO( r )$ current algebra.

The stochastic process is interpreted as the time evolution in a string field theory. The corresponding non-critical non-orientable open-closed string field theory is defined by the F-P hamiltonian operator. 
In terms of the expectation value of an observable $O(\phi, \psi)$, a
function of $\phi_n$'s and $\psi^{ab}_n$'s, the F-P hamiltonian operator ${\hat H}_{FP}$ is
defined by\cite{Na}
\EQ
<\phi (0), \psi (0)| {\rm e}^{- \t {\hat H}_{FP} } O({\hat \phi}, {\hat \psi})|0>
\equiv <O( \phi_{\xi\eta}(\t), \psi_{\xi\eta}(\t) )>_{\xi\eta}                  \  .
\EN
In R.H.S., $\phi_{\xi\eta}(\t)$ and $\psi_{\xi\eta}(\t)$ denote the solutions of the Langevin equations (4) 
with the initial configuration $\phi_n (0)$ and $\psi^{ab}_n (0)$. 
The F-P hamiltonian is equivalent to the differential operator in the well-known Fokker-Planck equation for the distribution functional except the operator ordering. In (7), ${\hat H}_{FP}$ is given by replacing the closed ( open ) string variable $\phi_n$ ( $\psi^{ab}_n$ ) to the creation operator ${\hat \phi}_n$ ( ${\hat \psi}^{ab}_n$ ) and the differential ${\pa \over \pa \phi_n}$ ( ${\pa \over \pa \psi^{ab}_n}$ ) to the annihilation operator ${\hat \pi}_n$ ( ${\hat \pi^{ab}_n}$ ), respectively. 


The continuum limit of ${\hat H}_{FP}$ is taken by introducing a length scale
\lq\lq $\ep$ " which defines the physical length of the strings created by
$\phi_n$ and $\psi^{ab}_n$ with
$
l \equiv n \ep
$. At the scaling limit, the string coupling constant, $G_{st}$, is kept finite with 
$
G_{st}
\equiv N^{-2} \ep^{- D}       \ .
$
The continuum stochastic time is defined by  
$
d\t
\equiv \ep^{- 2 + D/2} \v\t       \
$
.
We also redefine field variables 
at the continuum limit as follows.
\bea
\Phi (l)
&\equiv& \ep^{- D/2 } \phi_n      \ , \nn
\Pi (l)
&\equiv& \ep^{- 1 + D/2 } \pi_n      \ ,\nn
\Psi^{ab} (l)
&\equiv& \ep^{1 - D/2 } \psi^{ab}_n      \ , \nn
\Pi^{ab} (l)
&\equiv& \ep^{- 2 + D/2 } \pi^{ab}_n      \ .
\ena
The commutation relations are assumed to be 
$
\big[ \Pi(l) ,  \Phi(l')  \big]
= \d (l - l')               \ 
$
and 
$
\big[ \Pi^{ab}(l) ,  \Phi^{cd}(l')  \big]
= {1\over 2}(\d^{ac}\d^{bd} + \d^{ad}\d^{bc} )\d (l - l')       \  
$
.
Then we obtain the continuum F-P hamiltonian, ${\cal H}_{FP}$, from
$H_{FP}$ at the continuum limit,
\bea
{}\nn 
&& {\cal H}_{FP}         \nn
= &-& {1\over 2}\int_0^{\infty}\!dl
\big\{ 2 G_{st}\int_0^{\infty}\!dl' \Phi(l+l')l'\Pi(l')l\Pi(l) 
+ \int_0^{l}\!dl' \Phi(l-l')\Phi(l') l\Pi(l)                   
+ \sqrt{G_{st}} l\Phi(l) l\Pi(l)  \nn
&{}& \quad 
+ \sqrt{G_{st}} \sum_{a} \Psi^{aa}(l) l\Pi(l)      
+ 4 G_{st}\int_0^{\infty}\!dl' \Psi^{ab} (l+l')  l'\Pi^{ab}(l') l\Pi(l) 
+ \rho(l)\Pi(l)       \big\}             \nn 
&-& {1\over 2}\int_0^{\infty}\!dl \int_0^{\infty}\!dl' \int_0^{l}\!dk \int_0^{l'}\!dk' \big\{ \big(
\sqrt{G_{st}} \Psi^{ad}(l+ l' - k -k')\Psi^{bc}(k +k')   \nn
&{}& \qquad + \sqrt{G_{st}} \Psi^{ac}(l+ l' - k -k')\Psi^{bd}(k +k') 
\big) \Pi^{ab}(l) \Pi^{cd}(l')    \big\}      \nn
&+& \int^{\infty}_0\!dl \big\{ \sqrt{G_{st}} l\Psi^{ab}(l) l\Pi^{ab}(l) 
+ 2\int_0^l\!dl' \Phi(l-l') l'\Psi^{ab}(l') \Pi^{ab}(l)     \nn
&+& \sum_c \Psi^{ac}(l-l')\Psi^{cb}(l') \Pi^{ab}(l)     
- \sigma^{ab}(l) \Pi^{ab}(l)    \big\}    \ .    \nn 
\ena
By the redefinition (8), the F-P hamiltonian (9) is uniquely fixed
at the continuum limit except the cosmological terms. 
To specify the explicit form of the cosmological terms, $\rho(l)$ and $\sigma^{ab}(l)$ in (9), we consider the scaling limit in the simplest model, 
$g_0 = -1/2,\ g_1 = g/2, g_2 = g_3 =... = 0$ and 
$\mu^a_0 = -1,\ \mu^a_1 = \mu^a, \mu^a_2 = \mu^a_3 =... = 0$ in (1),
which corresponds to 2-dimensional pure gravity, $c=0$. 

The double scaling limit of the real symmetric matrix-vector model belongs to the 
same universality class as that of the real symmetric matrix model\cite{RS}. We notice that the fine tuning of the scale parameter $\lambda^a$ is necessary at the limit. The scaling limit is defined by, 
$
g 
= g_c {\rm e}^{ - c_1 \ep^2  t }   \ 
$
,
$
\mu^a 
= \mu^a_c {\rm e}^{ - m_a \ep }      \ 
$
and 
$
\lambda^a 
= { g\over 4\mu^a}    \ 
$
. Then these cosmological terms are found to be 
\bea
\rho(l) &=& 3 \d''(l) - {3 t\over 4}\d (l)    \ , \nn
\sigma^{ab}(l) &=& \d^{ab}\big( m^a \d(l) + 2\d' (l) \big)   \ . \nn
\ena
For $r=1$, (10) is consistent with those appeared in the orientable string field 
theory\cite{Mo}. 

The continuum F-P hamiltonian (9) takes the form of a linear combination of three continuum generators of string deformation, 
\bea
{\cal H}_{FP} 
&=& \int^{\infty}_0 dl \Big[ 
 G_{st} {\cal L}(l) l \Pi (l)          
+ \sqrt{G_{st}} \Big\{ 
{\cal K}^{ab} (l)      \nn
&{}& \qquad - {1\over4}\sum_c \int^l_0 dl' {l - 2l'\over l} \big( 
{\cal J}^{bc}(l-l')\Psi^{ac}(l) + {\cal J}^{ac}(l-l')\Psi^{bc}(l) 
\big)
\Big\}l \Pi^{ab}(l)      \Big]  \ .   \nn
\ena
The explicit forms of these continuum generators are given by 
\bea
{\cal L}(l)
&=& -
\big\{ \int_0^{\infty}\!dl' \Phi(l+l')l'\Pi(l') +
{1\over 2 G_{st}}\int_0^{l}\!dl' \Phi(l-l')\Phi(l')                    
+ \int_0^{\infty}\!dl' \Psi^{ab}(l+l')l'\Pi^{ab}(l')         \nn
&{}& \quad + {1\over 2\sqrt{G_{st}}}\sum_a \Psi^{aa} (l)               
+ {1\over 2 \sqrt{G_{st}}} l\Phi(l)
 + {1\over 2 G_{st}}{\rho(l)\over l}       \big\}             \ , \nn
{\cal K}^{ab} (l) 
&=& -  \big\{ 
\sqrt{ G_{st}}\int_0^{\infty}\!dl' \Psi^{ab} (l+l')  l'\Pi(l')     
+ {1\over 2} l\Psi^{ab}(l)  
+ {1\over 2\sqrt{G_{st}}}\int_0^l\!dl' \Phi(l-l') \Psi^{ab}(l')      \nn
&{}& \quad + {1\over 2}\sum_{cd}\int_0^{\infty}\!dl' \int_0^{l'}\!dk \Big(
\Psi^{ad}(l+ l' - k )\Psi^{bc}(k )   
+ \Psi^{bd}(l+ l' - k )\Psi^{ac}(k) 
\Big)  \Pi^{cd}(l')        \nn
&{}& \quad + {1\over 4\sqrt{G_{st}}}\sum_c \Big( 
\Psi^{ac}(l)\Psi^{cb}(0) + \Psi^{ac}(0)\Psi^{cb}(l)                \Big)  
- {1\over 2\sqrt{G_{st}}} {\sigma^{ab} \over l}
\big\}  \ , \nn
{\cal J}^{ab} (l) 
&=& - \big\{  
2\int^{\infty}_0 dl' \sum_c \Psi^{ac}(l+l')\Pi^{cb}(l') + {1\over
  \sqrt{G_{st}}}\d^{ab}\Phi(l)      
- {1\over
  \sqrt{G_{st}}}\big( m^b - {\pa\over \pa l}
\big)\Psi^{ab} (l)
\big\}              \ . \nn
\ena

The physical meaning of these generators are as follows. ${\cal L}(l)$, Virasoro generator, extends the string  with the length \lq\lq $l'$ \rq\rq to one with the length \lq\lq $l+l'$ \rq\rq, 
${\cal J}^{ab} (l)$ extends only the open string by changing its endpoints with the index \lq\lq b \rq\rq to \lq\lq a \rq\rq, 
${\cal K}^{ab} (l)$ extends both closed and open strings by cutting them and assigns the indices \lq\lq a \rq\rq and \lq\lq b \rq\rq to the open string new endpoints. In the sense of Ito calculus, the appearance of these generators can be traced to the noise correlations. Namely, $\v \zeta_{n-1}$ in (5) generates the deformation which corresponds to ${\cal L}(l)$ at the scaling limit. Similarly, $ \v \zeta^{ab}_{n-1}$ corresponds to ${\cal K}^{ab} (l)$ and $\v \eta^{ab}_{n}$, to ${\cal J}^{ab} (l)$. The remaining terms in these generators come from the matrix-vector model potential and the anomaly\cite{Na2}, which is equivalent to the Jacobian factor in the loop space path-integral measure.

These generators satisfy the algebra including the Virasoro algebra and the $SO( r )$ current algebra,
\bea
\big[{\cal L}(l), {\cal L}(l') \big]
&=& (l-l'){\cal L} (l+l')      \ , \nn
\big[ {\cal L} (l),  {\cal K}^{ab} (l') \big]  
&=& (l - l') {\cal K}^{ab}(l+l')         \nn 
&+& {1\over 4}\int_0^l\!dk \big\{ 
k \big( {\cal J}^{ac}(l-k) \Psi^{bc}(l'+k) + {\cal J}^{bc}(l-k)
\Psi^{ac}(l'+k)  \big)  \nn 
&-& k \big(  {\cal J}^{ac}(l'+k) \Psi^{bc}(l-k) + {\cal J}^{bc}(l'+k) \Psi^{ac}(l-k) \big)
\big\}    \ , \nn
\big[ {\cal L} (l),  {\cal J}^{ab} (l') \big] 
&=& - l' {\cal J}^{ab}(l+l')      \ , \nn
\big[ {\cal J}^{ab} (l),  {\cal K}^{cd}(l') \big] 
&=& - \d^{bd}{\cal K}^{ac}(l+l') - \d^{bc}{\cal K}^{ad}(l+l')         \nn
&+& {1\over 4}\d^{bd}\int_0^l\!dk \big( {\cal J}^{ce}(l+l'-k)\Psi^{ae}(k) 
-   {\cal J}^{ae}(l+l'-k)\Psi^{ce}(k) \big)    \nn
&+& {1\over 4}\d^{bc}\int_0^l\!dk \big( {\cal J}^{de}(l+l'-k)\Psi^{ae}(k) 
-   {\cal J}^{ae}(l+l'-k)\Psi^{de}(k) \big)            \nn
&+& {1\over 4}\int_0^l\!dk \big( {\cal J}^{db}(l+l'-k)\Psi^{ac}(k) 
+   {\cal J}^{cb}(k)\Psi^{ad}(l+l'-k)  \nn
&{}& \qquad + {\cal J}^{cb}(l+l'-k)\Psi^{ad}(k)
+ {\cal  J}^{bd}(k)\Psi^{ac}(l+l'-k) \big)             \ , \nn 
\big[ {\cal J}^{ab} (l),  {\cal J}^{cd} (l') \big] 
&=& - \d^{bc}{\cal J}^{ad}_{l+l'} + \d^{ad}{\cal J}^{cb}_{l+l'}    \ , \nn
\big[ {\cal K}^{ab} (l),  {\cal K}^{cd} (l') \big] 
&=& {1\over 4} \big( \int_0^l\!dk - \int_0^{l'}\!dk \big ) \big\{ 
 {\cal K}^{ac}(l+l'-k) \Psi^{bd}(k) +  {\cal K}^{ad}(l+l'-k)
 \Psi^{bc}(k) \nn
&{}& \qquad + {\cal K}^{bc}(l+l'-k) \Psi^{ad}(k) +  {\cal K}^{bd}(l+l'-k)
\Psi^{ac}(k)  
 \big\}    \nn
&+& {1\over 16}\int_0^{l+l'}\!dk \int_0^{l+l'-k}\!dk' \big\{  
{\cal J}^{ce}(l+l'-k-k')\big( \Psi^{ad}(k) \Psi^{be}(k') \nn 
&{}& \qquad + \Psi^{bd}(k) \Psi^{ae}(k') \big) 
+ \big( c \leftrightarrow d  \big)        \nn 
&-& {\cal J}^{ae}(l+l'-k-k')\big( \Psi^{bd}(k) \Psi^{ce}(k') + \Psi^{bc}(k) \Psi^{de}(k') \big) - \big( a \leftrightarrow b \big) \big\}         \ .    \nn
\ena
In the precise sense, these commutation relations are not an algebra because the open string creation operator $\psi^{ab}(l)$ appears in the R.H.S. of (13). The algebraic structure (13) is  understood as the consistency condition for the constraints or \lq\lq integrable condition \rq\rq. The symmetric part of the current, ${\cal J}^{ab}(l) + {\cal J}^{ba}(l) $, generates the S-D equation 
which is equivalent to the $\lambda$-independence of the equilibrium distribution. While the anti-symmetric part of the current, ${\cal J}^{ab}(l) - {\cal J}^{ba}(l)$, satisfies $SO( r )$ current algebra. 
We have taken the continuum limit in the simplest case $( c = 0)$. The extension of our analysis to $0<c<1$ cases is straightforward if we consider the multi-critical points of the matrix model\cite{NE}. 


In conclusion, we have derived the 
F-P hamiltonian which defines the non-critical non-orientable  
open-closed string field theory by applying SQM to real symmetric
matrix-vector models. The origin of the algebraic structure (13) can be traced to the noise correlations which generate the time evolution in the Langevin equation. Namely, the noise correlations realize the deformation of open and closed strings which are equivalent to those generated by the three constraints, ${\cal L}(l)$, ${\cal J}^{ab}(l)$ and ${\cal K}^{ab}(l)$. The appearance of the $SO(r)$ current algebra is consistent to the fact that the Chan-Paton realization of gauge groups in non-orientable strings is restricted to $SO(r)$ or $USp(r)$ due to the consistency of the string scattering amplitudes. Since the algebraic structure holds even in the case $0<c<1$, we expect the structure to be universal representing the scale invariance in the non-orientable open-closed string theories. The Langevin equation in Ito's calculus may provides a possible basis for numerical study of non-orientable open-closed string theories. We also hope that the structure will be useful to construct the type I superstring matrix model.


\end{document}